\documentclass[twocolumn,aps,prd,10pt,superscriptaddress,longbibliography,floatfix,citeautoscript]{revtex4-2}

\usepackage{amsmath}
\usepackage{amssymb}
\usepackage{bm}
\usepackage{dcolumn}
\usepackage{glossaries}
\usepackage{graphicx}
\usepackage{multirow}
\usepackage{fancyvrb}
\usepackage[linesnumbered,ruled]{algorithm2e}
\usepackage{siunitx}

\usepackage[usenames,dvipsnames,svgnames]{xcolor}
\usepackage{hyperref}
\hypersetup{
    pdfnewwindow=true,      
    colorlinks=true,        
    linkcolor=Blue,         
    citecolor=Blue,         
    filecolor=Blue,         
    urlcolor=Blue           
}

\usepackage{listings}	
\setacronymstyle{long-short}
\newacronym{dft}{DFT}{density functional theory}
\newacronym{dos}{DOS}{density of states}
\newacronym{gal}{GAL}{graphene antidot lattice}
\newacronym{gpu}{GPU}{graphics processing units}
\newacronym{pbe}{PBE}{Perdew-Burke-Ernzerhof}
\newacronym{lsqt}{LSQT}{linear-scaling quantum transport}
\newacronym{md}{MD}{molecular dynamics}
\newacronym{mbd}{MBD}{many-body dispersion}
\newacronym{mlp}{MLP}{machine-learned potential}
\newacronym{mlmd}{MLMD}{machine-learning molecular dynamics}
\newacronym{nep}{NEP}{neuroevolution potential}
\newacronym{rmse}{RMSE}{root mean square error}
\newacronym{snes}{SNES}{separable natural evolution strategy}
\newacronym{vdw}{vdW}{van-der-Waals}
\newacronym{2d}{2D}{two-dimensional}
\newacronym{hnemd}{HNEMD}{homogeneous non-equilibrium molecular dynamics}
\newacronym{mof}{MOF}{metal-organic framework}
\newacronym{tb}{TB}{tight-binding}
\newacronym{tdf}{TDF}{transport distribution function}

\DeclareSIUnit\angstrom{\text{Å}}
\DeclareSIUnit{\atom}{atom}
\DeclareSIUnit{\step}{step}
\DeclareSIUnit{\atomstepsecond}{\atom\step\per\second}

\newcolumntype{d}{D{.}{.}{-1}}

\begin{document}

\title{Combining linear-scaling quantum transport and machine-learning molecular dynamics to study thermal and electronic transports in complex materials}

\author{Zheyong Fan}
\email{brucenju@gmail.com}
\affiliation{College of Physical Science and Technology, Bohai University, Jinzhou 121013, P. R. China}

\author{Yang Xiao}
\affiliation{College of Physical Science and Technology, Bohai University, Jinzhou 121013, P. R. China}

\author{Yanzhou Wang}
\affiliation{MSP group, QTF Centre of Excellence, Department of Applied Physics, Aalto University, FI-00076 Aalto, Espoo, Finland}

\author{Penghua Ying}
\email{hityingph@163.com}
\affiliation{Department of Physical Chemistry, School of Chemistry, Tel Aviv University, Tel Aviv, 6997801, Israel}

\author{Shunda Chen}
\email{phychensd@gmail.com}
\affiliation{Department of Civil and Environmental Engineering, George Washington University,
Washington, DC 20052, USA}

\author{Haikuan Dong}
\email{donghaikuan@163.com}
\affiliation{College of Physical Science and Technology, Bohai University, Jinzhou 121013, P. R. China}

\date{\today}

\begin{abstract}
We propose an efficient approach for simultaneous prediction of thermal and electronic transport properties in complex materials. Firstly, a highly efficient machine-learned neuroevolution potential is trained using reference data from quantum-mechanical density-functional theory calculations. This trained potential is then applied in large-scale molecular dynamics simulations, enabling the generation of realistic structures and accurate characterization of thermal transport properties. In addition, molecular dynamics simulations of atoms and linear-scaling quantum transport calculations of electrons are coupled to account for the electron-phonon scattering and other disorders that affect the charge carriers governing the electronic transport properties. We demonstrate the usefulness of this unified approach by studying thermoelectric transport properties of a graphene antidot lattice.
\end{abstract}

\maketitle

\section{Introduction}

Thermal and electronic transports are two fundamental properties of a material. For simple solids, computational methods based on the electron and phonon Boltzmann transport equations \cite{ziman2001electrons} have been widely used to compute the transport properties mediated by the heat and charge carriers. There are a handful computational programs available for doing these calculations, such as \textsc{shengbte} \cite{li2014cpc}, \textsc{phono3py} \cite{Togo2015prb}, \textsc{kaldo} \cite{Barbalinardo2020jap}, and \textsc{gpupbte} \cite{Zhang2021jpcm} for thermal transport and \textsc{epw} \cite{ponce2016cpc}, \textsc{perturbo} \cite{zhou2021cpc}, and \textsc{phoebe} \cite{Cepellotti2022jpcm} for electronic transport. However, these methods can only efficiently deal with relatively simple systems and are generally not applicable to complex systems that cannot be properly represented by small periodic supercells.

To efficiently compute transport properties in complex systems one must resort to linear-scaling methods, i.e., methods with the computational cost that scales linearly with respect to the number of atoms in the periodic supercell. For thermal transport, \gls{md} simulation is such a linear-scaling method at the atomistic level \cite{gu2021jap}, provided that the interatomic potential used is a classical one and has a finite cutoff. Nowadays, \glspl{mlp} \cite{behler2016jcp} have been routinely applied in \gls{md} simulations of thermal transport. Particularly, the \gls{nep} \cite{fan2021prb,fan2022jpcm,fan2022jcp} has been developed with a focus on thermal transport applications and has excellent computational efficiency.

For electronic transport, there are also \gls{lsqt} methods \cite{fan2020pr} based on semi-empirical \gls{tb} models. The electron-phonon coupling in \gls{lsqt} calculations can be captured by the bond-length dependent hopping integrals in the electron \gls{tb} Hamiltonian \cite{Goringe1997rpp}. This has been explored using either specific phonon dynamics \cite{roche2005prl,roche2005prb} or \gls{md} simulations \cite{ishii2009crp,ishii2010prb,ishii2010prl}. Static-disorder approximation of the electron-phonon coupling has also been used for organic crystals \cite{ortmann2011prb,ciuchi2011prb}, graphene \cite{fan2017_2dm} and a carbon nanotube \cite{fan2018cpc}. Among these, the combined \gls{md}-\gls{lsqt} approach is the most flexible one, but it has not been widely used. The major reason is that there has been no accurate interatomic potential to drive \gls{md} simulations for a general system. Another reason is that there is so far no publicly available implementation of this approach.

In this paper, we propose to combine \gls{mlmd}, namely, \gls{md} driven by a \gls{mlp}, and \gls{lsqt} with a bond-length-aware \gls{tb} model, to study the thermal, electronic, and thermoelectric transport properties of complex materials that are beyond the reach of conventional methods. We call the combined method \gls{mlmd}-\gls{lsqt}. For the \gls{mlp}, we choose to use the highly efficient \gls{nep} approach \cite{fan2021prb,fan2022jpcm,fan2022jcp} as implemented in the \textit{open-source} graphics processing units molecular dynamics (\textsc{gpumd}) package \cite{fan2017cpc}. By training against quantum-mechanical \gls{dft} data, a \gls{nep} model can be constructed \textit{on demand}, which can then be used to perform large-scale \gls{md} simulations to obtain realistic structures and thermal transport properties. For the \gls{lsqt} part, we also implement it into the \textsc{gpumd} package (version 3.9) to couple electron and ion motions. To show the usefulness of this unified approach, we construct a general-purpose \gls{nep} for carbon systems and study thermal, electronic, and thermoelectric transport properties of patterned graphene that has large-scale structural features. 

\section{The MLMD-LSQT approach}

At the core of our method is the \gls{nep} approach \cite{fan2021prb,fan2022jpcm,fan2022jcp} for \gls{mlp} construction. It uses Chebyshev and Legendre polynomials to construct a local atom-environment descriptor of a given atom which is then mapped to the site energy $U_i$ of this atom via a feed-forward neural network. The free parameters in the neural network as well as the descriptor are optimized though the minimization of a loss function using an  evolutionary algorithm. The loss function is defined as a weighted sum of the \glspl{rmse} of energy, force, and virial between predictions and \gls{dft} target results in combination with regularization terms. This method as implemeted in \textsc{gpumd} \cite{fan2017cpc} has been shown to be able to achieve simultaneously the accuracy of \gls{dft} calculations and the computational cost of empirical potentials, allowing for large-scale \gls{md} simulations up to 8.1 million atoms using a single 40-gigabyte \gls{gpu} card \cite{liu2023prb}. 

The \gls{lsqt} method can be used to calculate electrical conductivity in large systems, but the prerequisite is to construct an electron Hamiltonian incorporating electron-phonon coupling and other disorders \cite{fan2020pr}. By using a bond-length-aware \gls{tb} model to configurations generated from \gls{md} simulations, electron-phonon coupling and other structural disorders can be effectively described. For dissipative electron transport, there are two equivalent ways to compute the electrical conductivity, one is based on the velocity-auto-correlation and the other is based on the mean-square displacement \cite{fan2020pr}. For the purpose of the present work, we found that the velocity-auto-correlation approach is more convenient because the time intervals used in the calculations are quite small, and the mean-square-displacement approach is only beneficial when the time intervals are large \cite{fan2014cpc}. 

In the velocity-auto-correlation approach, the electrical conductivity at energy $E$ and correlation time $t$ can be calculated as an integral
\begin{equation}
\label{equation:sigma}
    \Sigma(E,t)=\frac{2e^2}{\Omega} \int_0^{t} \mathrm{Tr} \left[\delta (E-\hat{H}) \mathrm{Re} (\hat{V}\hat{V}(\tau)) \right] d\tau,
\end{equation}
where $e$ is the elementary charge, $\Omega$ is the system volume, $\hat{H}$ is the electron Hamiltonian operator, $\delta(E-\hat{H})$ is the energy resolution operator, $\hat{V}$ is the velocity operator, and $\hat{V}(\tau)=e^{i\hat{H}\tau} \hat{V} e^{-i\hat{H}\tau}$ is the time-evolved velocity operator. To facilitate the discussion, we denote the trace in the integral as $C(E,\tau)$. The coupled \gls{mlmd}-\gls{lsqt} algorithm can be represented as follows:

\begin{enumerate}
\item Starting from an initial structure, run \gls{mlmd} for a number of steps in the isothermal or isothermal-isobaric ensemble to achieve equilibrium. 
\item Perform \gls{mlmd} simulation for a number of steps:
\begin{enumerate}
    \item  Evolve the atomic system from step $n-1$, $\{\bm{r}_i(n\Delta t - \Delta t)\}$, to step $n$, $\{\bm{r}_i(n\Delta t)\}$, by a time step of $\Delta t$ according the \gls{nep} interatomic potential. 
    \item Calculate the electron Hamiltonian and velocity operators at step $n$ according to the atom positions $\{\bm{r}_i(n\Delta t)\}$.
    \item Calculate $C(E,n\Delta t)$  using the electron Hamiltonian at the current step. In this step, linear-scaling techniques \cite{fan2020pr}, including sparse matrix-vector multiplication, random phase approximation of trace, Chebyshev expansion of quantum evolution operator, and kernel polynomial method \cite{weisse2006rmp} for energy resolution operator, are used.
\end{enumerate}
\end{enumerate}

After obtaining $C(E,\tau)$ at a number of discrete time points, it can be numerically integrated to calculate the electrical conductivity according to \autoref{equation:sigma}. This approach was implemented into the \textsc{gpumd} package and was available starting from version 3.9. Besides, the electronic \gls{dos} was also implemented according to the following expression:
\begin{equation}
\label{equation:rho}
    \rho(E)=\frac{2}{\Omega}  \mathrm{Tr} \left[\delta (E-\hat{H})  \right].
\end{equation}

\begin{figure*}[ht]
\centering
\includegraphics[width=2\columnwidth]{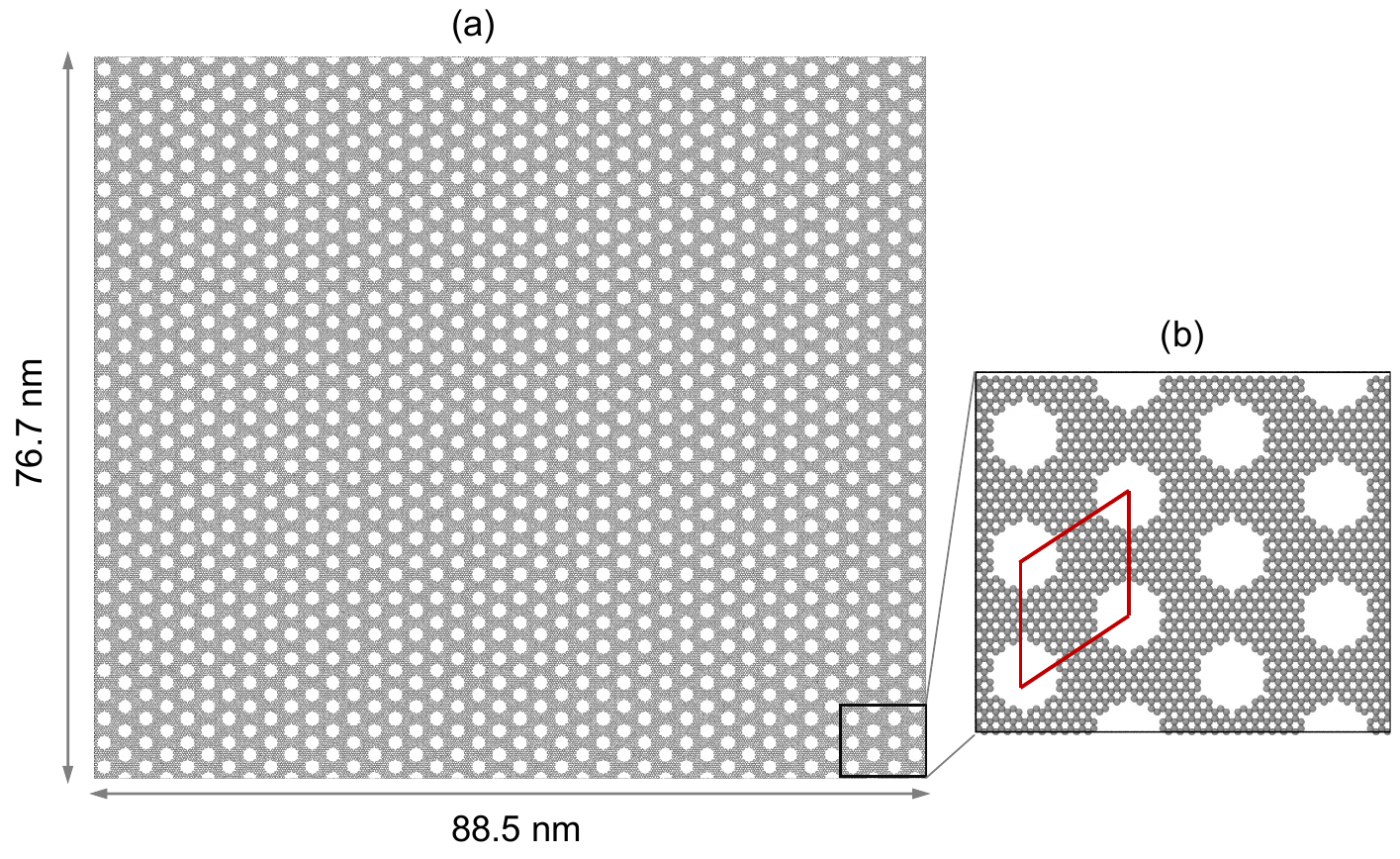}
\caption{(a) Atomistic structure of an example graphene antidot lattice (\gls{gal}) system studied in this work. (b) Illustration of the primitive cell containing 156 atoms, enclosed by the parallelogram.}
\label{fig:gal}
\end{figure*}

\section{Case study of a graphene antidot lattice}

As a proof of concept, we apply the \gls{mlmd}-\gls{lsqt} approach to study the thermoelectric transport in a \gls{gal} \cite{Pedersen2008prl}, also known as graphene nanomesh \cite{bai2010nnt}, a graphene sheet with patterned holes. 
Thermoelectric effects in graphene nanostructures have been extensively studied, and \glspl{gal} have been identified as one of the promising candidates for good thermoelectric materials
\cite{Dollfus2015jpcm}. However, previous works have only studied the ballistic electronic transport regime \cite{Gunst2011prb,Karamitaheri2011jap,yan2012pla}, without considering finite-temperature effects.

\autoref{fig:gal} shows the atomistic structure of the system under investigation. The simulation domain cell of the \gls{gal} sample contains \num{187200} atoms and has a dimension of about \SI{88.5}{\nano\meter} $\times$ \SI{76.7}{\nano\meter} in the $xy$-plane, which can be considered as a \gls{2d} system when periodic boundary conditions are applied to the in-plane directions. The thickness of the system was taken as \SI{0.335}{nm} in calculating the volume. The primitive cell for the \gls{gal} contains 156 atoms, a complexity that challenges conventional numerical methods based on the electron and phonon Boltzmann transport equations. However, this kind of complex structures are well-suited for the \gls{mlmd}-\gls{lsqt} approach. To construct the Hamiltonian and velocity operators, we employed a $p_z$-orbital \gls{tb} model with a bond-length dependent hopping parameter 
\begin{equation}
    H_{ij} = t_0 \left( \frac{r_0}{r_{ij}}\right)^2,
\end{equation}
where $t_0=$ \SI{-2.7}{\eV}, $r_0=$ \SI{1.42}{\angstrom}, and $r_{ij}$ is the distance between the atom pair $i$ and $j$. The model with a fixed hopping parameter $t_0$ has been used in previous works \cite{Pedersen2014prb,fan2015prb_electron} that did not account for electron-phonon coupling. The real-space Hamiltonian and velocity (assuming to be in the $x$ direction) operators can be written as
\begin{equation}
    \hat{H} = \sum_{i,j} H_{ij} |i \rangle\langle j|;
\end{equation}
\begin{equation}
    \hat{V} = \frac{i}{\hbar}\sum_{i,j} (x_j-x_i)H_{ij} |i \rangle\langle j|,
\end{equation}
where $x_i$ is the $x$-position of atom $i$.

\subsection{Training a general-purpose NEP for carbon systems}

\begin{figure}[ht]
\centering
\includegraphics[width=\columnwidth]{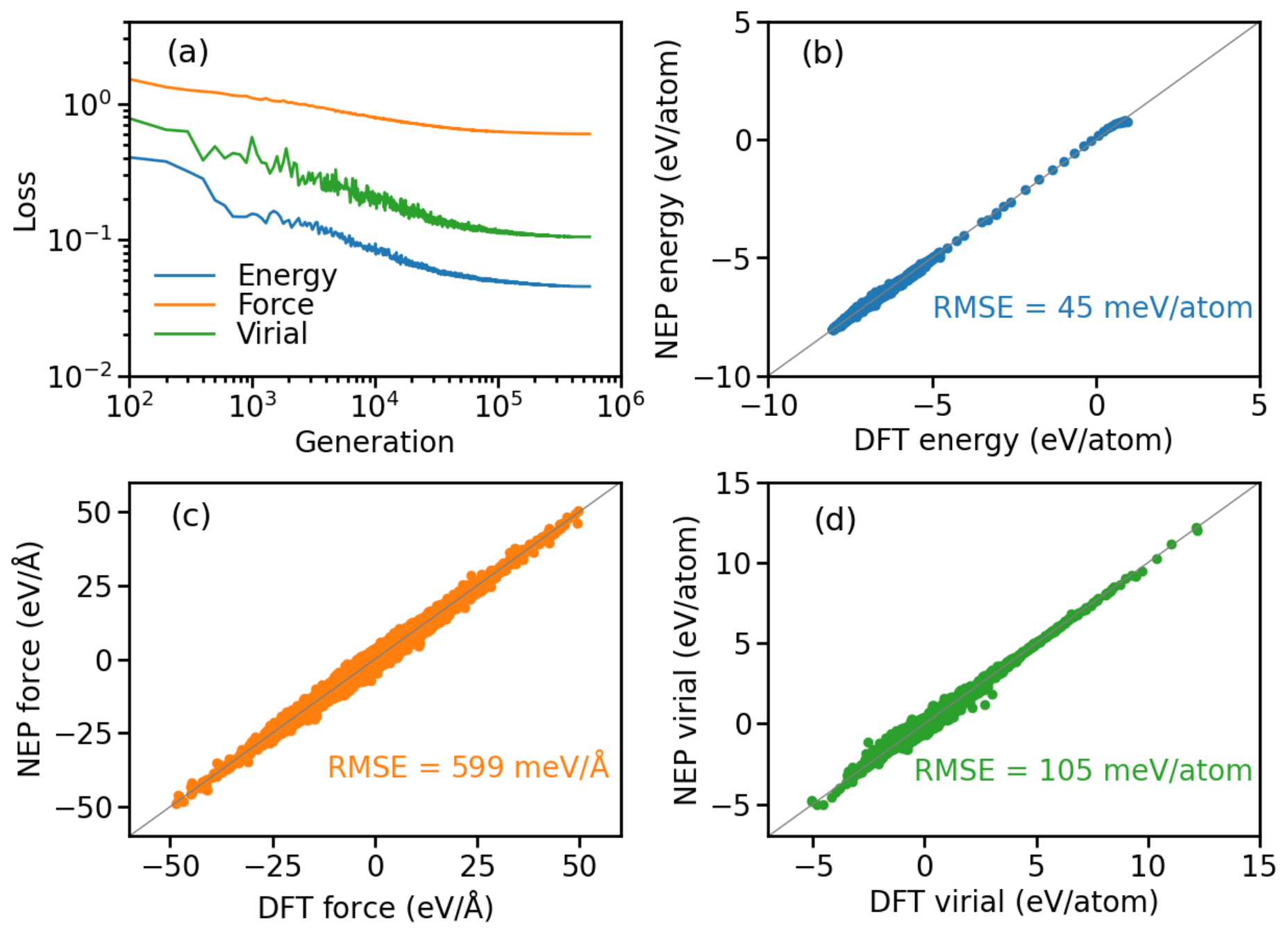}
\caption{(a) Evolution of the energy, force, and virial loss values as a function of the number of training generations for a general-purpose NEP for carbon systems  based on an extensive dataset \cite{Rowe2020jcp}. (b)-(d) Comparison between NEP predictions and DFT reference values for energy, force, and virial. The RMSE values are indicated in each panel. }
\label{fig:loss}
\end{figure}

Although for the scope of the current work, it suffices to train a specialized \gls{nep} model for \gls{gal}, it is our broader objective to train a general-purpose carbon potential based on the extensive dataset as used for constructing a Gaussian approximation potential \cite{Rowe2020jcp}. Using this dataset and the hyperparameters given in Appendix \ref{sec:nep_in}, we trained a general-purpose \gls{nep} model for carbon systems. The training results are shown in \autoref{fig:loss}. After a few hundred thousand training steps, the \glspl{rmse} of energy, force, and virial all converge [\autoref{fig:loss}(a)], and their converged values are 45 meV/atom, 599 meV/\AA{}, and 105 meV/atom, respectively. The predicted data are compared to the \gls{dft} reference ones in \autoref{fig:loss}(b)-(d). The seemingly large \gls{rmse} values are typical for general-purpose carbon systems, as similar ones were reported in or can be extracted from previous works \cite{Rowe2020jcp,fan2022jcp,wang2022carbon,wang2022cm,Qamar2023jctc}.

\subsection{Thermal transport}

\begin{figure}[htb]
\centering
\includegraphics[width=\columnwidth]{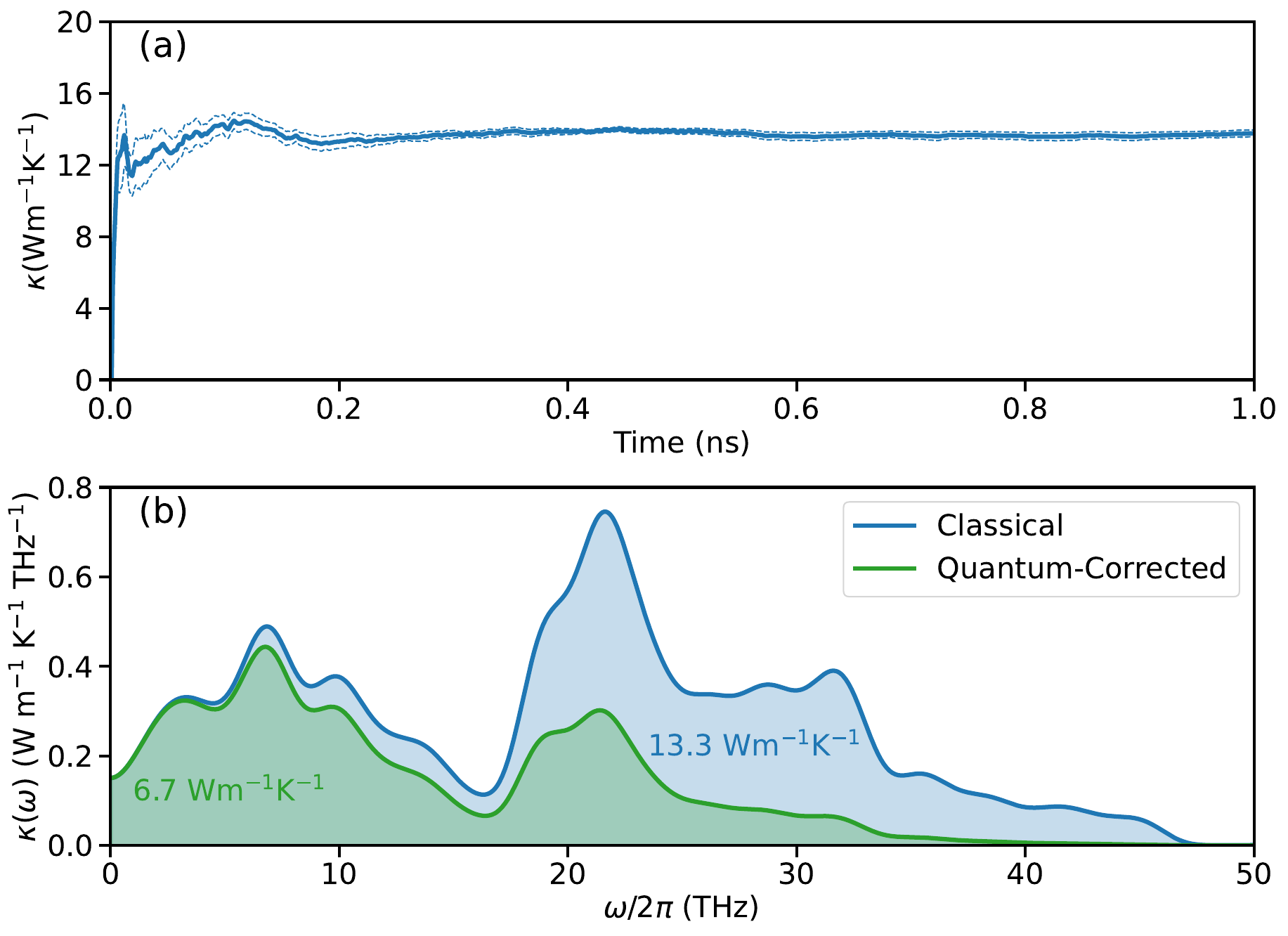}
\caption{(a) Phonon thermal conductivity $\kappa_{\rm ph}$ as a function of the \gls{hnemd} production time for GAL at 300 K. (b) The classical and quantum-corrected spectral thermal conductivity as a function of the phonon frequency $\omega/2\pi$.}
\label{fig:kappa}
\end{figure}

For a complete study of thermoelectric transport, the lattice (phonon) thermal conductivity $\kappa_{\rm ph}$ must be evaluated. To this end, we calculated $\kappa_{\rm ph}$ for the \gls{gal} model with \num{187200} atoms using the \gls{hnemd} method \cite{fan2019prb}. In this method, an external driving force
\begin{equation}
\bm{F}_i^{\mathrm{ext}}=\bm{F}_{\mathrm{e}} \cdot \sum_{j \neq i}\left(\frac{\partial U_j}{\partial \bm{r}_{j i}} \otimes \bm{r}_{i j}\right)
\end{equation}
is exerted on each atom $i$, driving the system out of equilibrium. Here, $\bm{F}_{\mathrm{e}}$ is the driving force parameter with the dimension of inverse length and $\bm{r}_{i j} \equiv \bm{r}_j-\bm{r}_i$, $\bm{r}_i$ being the position of atom $i$. After a steady state is achieved, the lattice thermal conductivity tensor $\kappa_{\rm ph}^{\alpha\beta}$ can be computed from the relation
\begin{equation}
\frac{\left\langle J^\alpha\right\rangle}{T \Omega}=\sum_\beta \kappa_{\rm ph}^{\alpha \beta} F_{\mathrm{e}}^\beta,
\end{equation}
where $T$ is the system temperature, $\Omega$ is the system volume, and $\langle J^{\alpha} \rangle$ is the ensemble average of the heat current \cite{fan2015prb}
\begin{equation}
    \bm{J} = \sum_i \bm{v}_i \cdot \sum_{j \neq i}\left(\frac{\partial U_j}{\partial \bm{r}_{j i}} \otimes \bm{r}_{i j}\right).
\end{equation}
In this case study, we only consider the condition of $300$ K and zero in-plane pressure. The input script for \textsc{gpumd} is given in Appendix \ref{sec:run_in_kappa}. The time convergence of $\kappa_{\rm ph}$ is shown in \autoref{fig:kappa}(a).
We have performed 4 independent simulations in both the $x$ and the $y$ directions and averaged  the results over the two directions as the system is essentially isotropic.

For the \gls{gal} in our case study, the simulation temperature (300 K) is much lower than the Debye temperature (on the order of 2000 K), and the classical \gls{md} simulation thus significantly overestimates the modal heat capacity, which in turn leads to an overestimation of the thermal conductivity. Fortunately, there exists a feasible correction for the missing quantum statistics, as has been successfully applied to amorphous \cite{wang2023prb,zhang2023prb} and fluid \cite{xu2023jcp} systems described by \gls{nep} models. In this quantum correction method, the spectral thermal conductivity $\kappa(\omega)$ as calculated from the \gls{hnemd} method \cite{fan2019prb} is multiplied by a quantum-to-classical factor $p(x)=x^2 e^x/\left(e^x-1\right)^2$, where $x=\hbar \omega / k_{\mathrm{B}} T$, $\omega$ is the phonon frequency, $\hbar$ is the reduced Planck constant, and $k_{\mathrm{B}}$ is the Boltzmann constant. The spectral decomposition of $\kappa_{\rm ph}$ corresponding to the classical results is depicted in \autoref{fig:kappa}(b), where the quantum-corrected results are also shown. The classical value of $\kappa_{\rm ph}$ is \SI{13.3}{\watt\per\meter\per\kelvin}, which becomes \SI{6.7}{\watt\per\meter\per\kelvin} after quantum correction. The quantum corrected value will be used later. 

\subsection{Electronic and thermoeletric transports}

\begin{figure}[ht]
\centering
\includegraphics[width=\columnwidth]{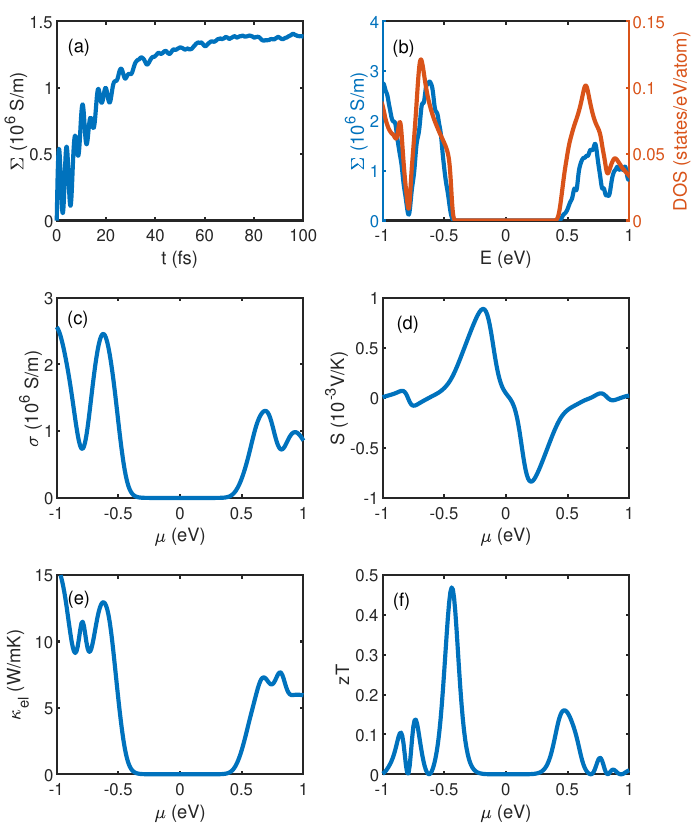}
\caption{(a) Electrical conductivity $\Sigma(E,t)$ as computed by using \autoref{equation:sigma} at $E=$ \SI{0.65}{\eV}. (b) Transport distribution function (\gls{tdf}) $\Sigma(E)$ and electronic density of states (\gls{dos}). (c)-(f) Electrical conductivity $\sigma(\mu,T)$, Seebeck coefficient $S(\mu,T)$, electronic thermal conductivity $\kappa_{\rm el}(\mu,T)$, and figure of merit $zT(\mu,T)$ for a range of chemical potential $\mu$ at $T=$ \SI{300}{\kelvin}. }
\label{fig:lsqt}
\end{figure}

The time-dependent electrical conductivity $\Sigma(E,t)$ (see Appendix \ref{sec:run_in_sigma} for the input script and more calculation details) converges in the diffusive transport regime, and one can obtain the so-called semi-classical electrical conductivity $\Sigma(E)$ by averaging $\Sigma(E,t)$ over a proper range of correlation time,
\begin{equation}
  \Sigma(E) = \frac{1}{t_2-t_1} \int_{t_1}^{t_2} \Sigma(E,t) dt.  
\end{equation}
According to \autoref{fig:lsqt}(a), it is a good choice to set $t_1=$ \SI{80}{\femto\second} and $t_2=$ \SI{100}{\femto\second}. The semi-classical electrical conductivity  $\Sigma(E)$ can be regarded as the \gls{tdf} \cite{Mahan1996pnas,fan2011jap,Zhou2011prl,Jeong2012jap,Maassen2021prb,ding2023npjcm} for thermoelectric transport. The calculated \gls{tdf} as well as \gls{dos} are presented in \autoref{fig:lsqt}(b). The anti-dots induce a considerable band gap of about 0.8 eV. This band gap is then also the transport gap. 

From the \gls{tdf}, we then calculated the transport coefficients at 300 K for a range of chemical potential $\mu$. We first define the \textit{functionals} ($n=0,1,2$) of the \gls{tdf}:
\begin{equation}
 X_n(\mu,T) = \int \left[-\frac{\partial f(E,\mu,T)}{\partial E}\right] E^n\Sigma(E) dE,
\end{equation}
where 
\begin{equation}
f(E,\mu,T) = \frac{1}{\exp\left(\frac{E-\mu}{k_{\rm B} T} \right)+1}
\end{equation} is the Fermi-Dirac distribution. The electrical conductivity $\sigma(\mu,T)$, Seebeck coefficient $S(\mu,T)$, and electronic thermal conductivity $\kappa_{\rm el}(\mu,T)$ can be expressed in terms of these functionals as
\begin{equation}
\sigma(\mu,T) = X_0(\mu,T);
\end{equation}
\begin{equation}
S(\mu,T) = -\frac{1}{eT} \left[ \frac{X_1(\mu,T)}{X_0(\mu,T)}-\mu\right];
\end{equation}
\begin{equation}
\kappa_{\rm el}(\mu,T) = \frac{1}{e^2T} \left[X_2(\mu,T) - \frac{X_1^2(\mu,T)}{X_0(\mu,T)}\right].
\end{equation}
The calculated results are presented in \autoref{fig:lsqt}(c)-(e). The finite-temperature electrical conductivity $\sigma(\mu,T)$ resembles the \gls{tdf}, but with smearing resulting from the Fermi-Dirac distribution. The Seebeck coefficient has a negative peak for electrons and a positive peak for holes. The electronic thermal conductivity resembles the electrical conductivity in shape which is in line with the Wiedemann-Franz law.  

Based on these transport coefficients and the (quantum-corrected) phonon thermal conductivity $\kappa_{\rm ph}$, one can define the dimensionless figure of merit as 
\begin{equation}
 zT(\mu,T) = \frac{S^2(\mu,T) \sigma(\mu,T)} { \kappa_{\rm ph} + \kappa_{\rm el}(\mu,T) } T.
\end{equation}
Due to the competition between the various transport coefficients, $zT$ develops peaks for both electron and hole transport, at $\mu=$ \SI{0.47}{\eV} and \SI{-0.44}{\eV}, respectively. The transport is asymmetric between electron and hole, showing a maximum $zT=0.47$ for hole and a maximum $zT=0.16$ for electron. 

Experimentally, thermoelectric transport properties have been measured for single- and bi-layer graphene nanomeshes with the neck width down to \SI{8}{\nm} \cite{Jinwoo2017ne}. This neck width between the nearest antidot pairs is a few time larger than that we studied and the measured thermal conductivity values (of the order of \SI{100}{\watt\per\meter\per\kelvin}) are significantly larger than our prediction. On the other hand, there are geometrical disorders in the experimental samples, namely, variations in the positions and sizes of the antidots, which, according to previous calculations \cite{Pedersen2014prb,fan2015prb_electron}, can lead to suppressed electrical conductivity. Therefore, the relatively high $zT$ values we predicted remain a challenge for experimental realization. 

\section{Summary and conclusions}

In summary, we have introduced a numerical approach for simultaneous prediction of thermal and electronic transport properties in complex materials. This approach, based on \gls{mlmd} and \gls{lsqt}, offers an excellent efficiency with a computational cost that scales linearly with the system size. For a given material, a highly efficient \gls{nep} is first constructed \textit{on demand}. This \gls{mlp} can be used to perform large-scale \gls{md} to obtain realistic structures and accurate thermal transport properties. By combining the time-evolution of electrons and atoms during the \gls{md} simulation, electron-phonon scattering and other disorders for the charge carriers can be naturally captured and the various electronic transport properties can be obtained. 

As an illustrative example, we have investigated the thermoelectric transport properties of a type of graphene antidot lattices (\glspl{gal}), predicting its relatively high thermoelectric efficiency at room temperature. We recognize the necessity of future work to conduct a more comprehensive study of the thermoelectric transport in \glspl{gal} using our proposed approach. Subsequent research endeavors may consider integrating machine-learning techniques to explore the vast design space, similar to the methods employed in thermal transport studies in these systems \cite{wan2020carbon,wei2020ne}. 

\begin{acknowledgments}
Z. Fan and H. Dong were supported by the National Natural Science Foundation of China (NSFC) (No. 11974059) and the Research Fund of Bohai University (No. 0522xn076). P. Ying was supported by the Israel Academy of Sciences and Humanities \& Council for Higher Education Excellence Fellowship Program for International Postdoctoral Researchers.  
\end{acknowledgments}

\vspace{0.5cm}
\noindent{\textbf{Data availability:}}

Complete input and output files for the general-purpose carbon \gls{nep} model are freely available at \url{https://gitlab.com/brucefan1983/nep-data}.
The source code and documentation for \textsc{gpumd} are available
at \url{https://github.com/brucefan1983/GPUMD} and \url{https://gpumd.org}, respectively.

\vspace{0.5cm}
\noindent{\textbf{Declaration of competing interest:}}

The authors declare that they have no competing interests.

\appendix

\section{Inputs for training the  NEP model}
\label{sec:nep_in}

\gls{nep} models can be trained using the \verb"nep" executable in the \textsc{gpumd} package. The relevant hyperparameters are specified in the \verb"nep.in" input file. The contents of the \verb"nep.in" input file for training the general-purpose model of carbon systems are given below. 
\begin{verbatim}
type         1 C
version      4
cutoff       7  4
n_max        12  8 
basis_size   16 12
l_max        4 2 1
neuron       100
lambda_1     0.0
lambda_e     1.0
lambda_v     0.1
batch        8000
population   100
generation   2000000  
\end{verbatim}

\section{Inputs for phonon thermal conductivity calculations}
\label{sec:run_in_kappa}

\gls{md} simulations with \gls{nep} models can be performed by using the \verb"gpumd" executable in the \textsc{gpumd} package. The controlling parameters are specified in the \verb"run.in" input file. The contents of the \verb"run.in" input file for calculating the thermal conductivity are given below. 

\begin{verbatim}
# setup
potential    nep.txt
velocity     300

# equilibration
ensemble     npt_ber 300 300 100 0 0 0 
             1000 1000 1000 1000 
time_step    1         
run          100000

# production
ensemble        nvt_nhc 300 300 100
compute_hnemd   1000 1e-4 0 0
compute_shc     2 250 0 1000 400.0 
run             1000000
\end{verbatim} 

\section{Inputs for electronic transport calculations}
\label{sec:run_in_sigma}

The contents of the \verb"run.in" input file for calculating the electronic transport properties are given below. The time step in the production stage is chosen to be small enough (0.1 fs) to ensure accurate integration in \autoref{equation:sigma}. The keyword \verb"compute_lsqt" invokes the \gls{lsqt} calculations. This is a new keyword introduced in GPUMD-v3.9 during the course of the present study. Here are the meanings of the parameters for this keyword:
\begin{itemize}
    \item The first parameter \verb"x" means that the transport is along the $x$ direction. We have calculated 10 times along the $x$ directions and also 10 times along the $y$ direction and averaged the results. 
    \item The second parameter refers to the number of Chebyshev moments in the kernel-polynomial method \cite{weisse2006rmp} for both the \gls{dos} and conductivity calculations. A value of \num{3000} is large enough here.
    \item The next three parameters are respectively the number of energy points to be considered, the minimum energy and the maximum energy. Here we calculated the transport properties from \SI{-8.1}{\eV} to \SI{8.1}{\eV}, with an interval of \SI{1.62}{\meV}. 
    \item The last parameter is an energy threshold that needs to be larger than the energy range of the tight-binding model. Here, a value of \SI{8.2}{\eV} is sufficient. This parameter can be determined by a trial-and-error approach.
\end{itemize}

\begin{verbatim}
#setup
potential       nep.txt
velocity        300

# equilibration
ensemble     npt_ber 300 300 100 0 0 0 
             1000 1000 1000 1000 
time_step    1
dump_exyz    100000
run          100000

# production
ensemble     nve
time_step    0.1
compute_lsqt x 3000 10001 -8.1 8.1 8.2
run          1000
\end{verbatim}

\end{document}